\documentstyle[12pt]{article}
\begin{document}
\author{\normalsize\bf E.I. Semenov
\thanks{e-mail: semenov@icc.ru}}
\title{Some properties of the equation of fast diffusion and its
multidimensional exact solutions\\}
\date{\it Institute of System Dynamics
and Control Theory\\
Siberian Branch
of Academy of Scienses of Russia,\\
P.O. Box 1233, 664033 Irkutsk, Russia}
\thispagestyle{empty}
\maketitle{}

\def\sh{\mathop{\hbox{sh}}\limits}
\def\ch{\mathop{\hbox{ch}}\limits}
\def\th{\mathop{\hbox{th}}\limits}
\def\cth{\mathop{\hbox{cth}}\limits}
\def\tg{\mathop{\hbox{tg}}\limits}
\def\sec{\mathop{\hbox{sec}}\limits}
\def\sech{\mathop{\hbox{sech}}\limits}
\def\Rel{\mathop{\hbox{Rel}}\limits}

\[
{\bf Abstract}
\]

The invariance for the equation of fast diffusion in the 2D coordinate
space has
been proved, and its reduction to the 1D (with respect to the
spatial variable) analog is demonstrated. On the basis of these
results, new exact multi-dimensional solutions, which are dependent
on arbitrary harmonic functions, are constructed. As a result, new
exact solutions of the well-known Liouville equation -- the
steady-state analog for the fast diffusion equation with the linear
source -- have been obtained. Some generalizations for the systems
of quasi-linear parabolic equations, as well as systems of elliptic
equations with Poisson interaction, which are applied in the theory
of semiconductors, are considered.

\begin{center}
\large{\bf Introduction}
\end{center}

The paper investigates the equation of fast diffusion
$$
u_t=\Delta\ln u, \;\; u=u(x,y,t),
\eqno(1)
$$
in the 2D coordinate space, which is characteristic of many
applied problems, for example, in the description of spreading
of superfine monomolecular layers of liquid under the
influence of Van der Waals forces [1]. It arises in modeling
of diffusion phenomena in semiconductors, polymers, etc. It is
known [2] that eq. (1) is special  from the viewpoint
of the group theory since it assumes finite-dimensional algebra of
point symmetries. This means that eq. (1) in the 2D coordinate space
possesses an infinite stock of invariant solutions [3]. In the
literature, the relation (1) is sometimes called the Ricci equation.

This paper considers some properties of (1). On the basis of these
properties some new -- exact and multidimensional -- solutions of
this equation, which are dependent on arbitrary harmonic functions,
are constructed.

\section{Invariance of the equation of fast diffusion}

Let invariance be understood as nonvariability of the form of
this equation under the effect of any transformation.

{\bf Defenition}[4]. $\;$
Harmonic functions $\xi (x,y)$ and $\eta (x,y)$ are called
conjugate in the simple connected domain $D$ if the function
$F(z)=\xi (x,y) + i\eta (x,y)$ is some analytical function of the
argument $z=x + iy$ in the domain $D$.

Conjugate harmonic functions are related by Cauchy-Riemann equations
$$
\frac{\partial\xi}{\partial x}=
\frac{\partial\eta}{\partial y},\;\;
\frac{\partial\xi}{\partial y}=-
\frac{\partial\eta}{\partial x},
$$
define one another everywhere in $D$ with the  to
an additive constant and, consequently, possess the following
properties
$$
(\nabla\xi,\nabla\eta)=0, \;\;  |\nabla\xi|^2=|\nabla\eta|^2.
\eqno(2)
$$
The harmonic polynomials
$$
\xi (x,y)=\left (x^2 + y^2\right )^{\frac{n}{2}}\cos(n\varphi),\;\;
\eta (x,y)=\left (x^2 + y^2\right )^{\frac{n}{2}}\sin(n\varphi),
$$
where
$$
\varphi=\arccos\left(\frac{x}{\sqrt{x^2 + y^2}}\right),\;\;  {\rm or}\;\;
\varphi=\arcsin\left(\frac{y}{\sqrt{x^2 + y^2}}\right),\;  n\in N,
$$
are the simplest examples of the conjugate harmonic functions.

{\bf Lemma 1.}$\;${\it
If the function $\eta (x,y)$ is harmonic then
the function $\displaystyle\ln\left |\eta_x^2 + \eta_y^2\right|$
is also harmonic.}

{\bf Proof.}
For the purpose of comfort let us introduce the denotation
$\psi (x,y)=\eta_x^2 + \eta_y^2.$
$$
\Delta\ln\psi=\left [\psi\left(\psi_{xx} + \psi_{yy}\right )-
\left (\psi_x^2 + \psi_y^2\right )\right ]\psi^{-2}.
\eqno(3)
$$
By consecutive computing necessary partial derivatives, we obtain
$$
\psi_{xx} + \psi_{yy}=2\left (\eta_{xx}^2 + \eta_{yy}^2\right ) + 4\eta_{xy}^2 +
2\eta_{x}\left (\eta_{xx} + \eta_{yy}\right )_x +
2\eta_{y}\left (\eta_{xx} + \eta_{yy}\right )_y,
$$
$$
\psi_{x}^2 + \psi_{y}^2=8\eta_{x}\eta_{y}\eta_{xy}
\left (\eta_{xx} + \eta_{yy}\right ) +
4\eta_{x}^2\left (\eta_{xx}^2 + \eta_{xy}^2\right ) +
4\eta_{y}^2\left (\eta_{yy}^2 + \eta_{xy}^2\right ).
$$
Since due to the harmonic character of $\eta_{xx}=-\eta_{yy},$
the latter relations will, respectively, assume the following form:
$$
\psi_{xx} + \psi_{yy}=4\left (\eta_{xx}^2 + \eta_{xy}^2\right ),
$$
$$
\psi_{x}^2 + \psi_{y}^2=4\left (\eta_{x}^2 + \eta_{y}^2\right )
\left (\eta_{xx}^2 + \eta_{xy}^2\right ).
$$
Hence, expression (3) turns identically zero. $\Box$

Now let us formulate one of the main results of the work.

{\bf Theorem 1.}$\;${\it
If $\xi=\xi (x,y),  \eta=\eta (x,y)$
are conjugate harmonic functions then the equation of fast diffusion (1)
is invariant with respect to the transformation
$$
u(x,y,t)=\rho (x,y)v(\xi,\eta,t),
\eqno(4)
$$
where $\rho (x,y)=|\nabla\eta|^2,$ i.e. under the effect of (4) it
transforms into itself
$$
v_t=\Delta_{\xi\eta}\ln v.
\eqno(5)
$$
}

{\bf Proof.}
After substitution of relation (4) into eq. (1)
and simple computations, on account of the properties of
conjugate harmonic functions (2), we obtain
$$
\rho v_t=\Delta_{xy}\ln\rho + \rho\Delta_{\xi\eta}\ln v.
$$
Hence, due to Lemma 1, relation (5) immediately follows. $\Box$

Therefore, expression (4) is a form of autotransformation for
the equation of fast diffusion (1) and, consequently, it is the
formula of branching solutions.

{\bf Example 1.}$\;$  Consider a exact solution of eq. (5) of the form [5, 6]
$$
v(\xi ,\eta ,t)=2\frac{{\th}(t)}{\xi^2 + \eta^2{\th}^2(t)}.
$$
By applying transformation (4) with arbitrarily chosen harmonic
functions to this solution let us construct a few new exact
nonautomodel explicit solutions of eq. (1), which are anisotropic
with respect to spatial variables
$$
u(x,y,t)=2\frac{\left({\cth}^2(x) +{\tg}^2(y)\right){\th}(t)}
{1 + {\cth}^2(x){\tg}^2(y){\th}^2(t)},
$$
$$
u(x,y,t)=2\frac{\left({\tg}^2(x) +{\th}^2(y)\right){\th}(t)}
{1 + {\tg}^2(x){\th}^2(y){\th}^2(t)},
$$
$$
u(x,y,t)=18\frac{\left ({\tg}^2\left (x^3 - 3xy^2\right ) +
{\th}^2\left(3x^2y - y^3\right )\right )\left (x^2 + y^2\right )^2{\th}(t)}
{1 + {\tg}^2\left (x^3 - 3xy^2\right ){\th}^2\left(3x^2y - y^3\right ){\th}^2(t)},
$$
$$
u(x,y,t)=18\frac{\left
({\tg}^2\left (\exp(3x)a(y)\right ) +
{\th}^2\left (\exp(3x)b(y)\right )
\right )\exp(x){\th}(t)}
{1 +
{\tg}^2\left (\exp(3x)a(y)\right )
{\th}^2\left(\exp(3x)b(y)\right ){\th}^2(t)},
$$
where the following denotations are used
$$
a(y)=\cos^3(y) - 3\cos(y)\sin^2(y),\;\;
b(y)=3\cos^2(y)\sin(y) - \sin^3(y).
$$
The solutions obtained are significant in virtue of the fact that for
$t\rightarrow +\infty$ they are stabilized in the form of steady-state
solutions.

A result  similar to that of Theorem 1 can be extended onto some class
of systems of quasilinear parabolic equations.  Hence,  the  following
theorem is valid.

{\bf Theorem 2.}$\;${\it
Let $\xi=\xi (x,y),\; \eta=\eta (x,y)$
are conjugate harmonic functions, and
$f_{i}(u_{1}, u_{2}, \ldots, u_{m}),  i=1,2,\ldots ,n,   m\leq n,$
are homogeneous, the degree of homogeneity being one, i.e.
$$
f_{i}(\lambda u_{1},\lambda u_{2}, \ldots,\lambda u_{m})=
\lambda f_{i}(u_{1}, u_{2}, \ldots, u_{m}).
\eqno(6)
$$
The n the system of $n$ equations
$$
\frac{\partial u_{i}}{\partial t}=
\Delta_{xy}\ln u_{i} + f_{i}(u_{1}, u_{2}, \ldots, u_{m})
\eqno(7)
$$
is invariant with respect to the transformations
$$
u_{i}(x,y,t)=\rho (x,y)v_{i}(\xi,\eta,t),
\eqno(8)
$$
where $\rho =|\nabla\eta|^2,$ i.e. it transforms into itself
$$
\frac{\partial v_{i}}{\partial t}=
\Delta_{\xi\eta}\ln v_{i} + f_{i}(v_{1}, v_{2}, \ldots, v_{m}).
\eqno(9)
$$
}

{\bf Proof.}
By substituting functions (8) into eqs. (7) and taking account of
the conditions of homogeneity (6), we obtain
the system
$$
\rho\frac{\partial v_{i}}{\partial t}=
\Delta_{xy}\left (\ln\rho + \ln v_{i}\right )
+ \rho f_{i}(v_{1}, v_{2}, \ldots, v_{m}),
\eqno(10)
$$
furthermore, when using the properties of the conjugate harmonic
functions (2), it can readily be shown that
$$
\Delta_{xy}\ln v_{i}=\rho\Delta_{\xi\eta}\ln v_{i}.
$$
Therefore, from (10), due to Lemma 1, it is possible to obtain
the system of equations (9). $\Box$

\section{Reduction of the equation of fast diffusion to its
one-dimensional (with respect to the spatial variable) analog}

{\bf Theorem 3.}$\;${\it
Let $\eta (x,y)$ be an arbitrary harmonic function, which is different
from the constant one. Then with the aid of the transformation
$$
u(x,y,t)=\left (\eta_{x}^2 + \eta_{y}^2\right )v(\eta,t),
\eqno(11)
$$
the equation of fast diffusion (1) can be reduced to its
one dimensional (with respect to the spatial variable $\eta$)
analog
$$
\frac{\partial v}{\partial t}=
\frac{\partial^2\ln v}{\partial\eta^2}.
\eqno(12)
$$
}

{\bf Proof.}
After substituting expression (11) into eq. (1), we obtain
the equality
$$
\psi (x,y)\frac{\partial v}{\partial t}=
\Delta_{xy}\left (\ln\psi (x,y) + \ln v\right ),
\eqno(13)
$$
furthermore, it can easily be shown that
$$
\Delta_{xy}\ln v=\psi (x,y)\frac{\partial^2\ln v}{\partial\eta^2},
$$
where $\psi (x,y)=\eta_{x}^2 + \eta_{y}^2.$ Therefore, from equality (13),
due to Lemma 1, we obtain eq. (12). $\Box$

Consequently, by integrating eq. (12) and using formula (11),
it is possible to construct a class of exact solutions of fast
diffusion equations (1), which depend on an arbitrary harmonic function.

{\bf Example 2.}$\;$ Let $\eta (x,y)$ be a 4th degree harmonic polynomial,
i.e. $\eta (x,y)=x^4 - 6x^2y^2 + y^4$,  or
$\eta (x,y)=4\left (x^3y - xy^3\right ).$ Hence
$\eta_{x}^2 + \eta_{y}^2=16\left (x^2 + y^2\right)^3.$ In this case,
the transformation
$$
u(x,y,t)=16\left (x^2 + y^2\right)^3v(\eta,t),
$$
gives an anisotropic (with respect to spatial variables $x$ and $y$)
solution of eq.  (1), furthermore, the function $v(\eta,t)$ can be
determined from the relation (12). Since with the aid of the
substitution $v=w^{-1}$ eq. (12) can be reduced to the equation with
quadratic nonlinearities
$$
w_{t}=ww_{\eta\eta} - \left (w_{\eta}\right )^2,
$$
it is possible, in addition, to construct some exact solutions
of eq. (12) on the basis of the approach described in [7].
$$
v(\eta, t)=\frac{1}{\lambda}\cdot
\frac{\sqrt{k_1^2 + k_2^2}{\sh}(\lambda t)}
{k_1\cos(\eta) + k_2\sin(\eta) + \sqrt{k_1^2+k_2^2}{\ch}(\lambda t)},
$$
$$
v(\eta, t)=\frac{1}{\lambda}\cdot
\frac{\sqrt{k_1^2 + k_2^2}\cos(\lambda t)}
{k_1\cos(\eta) + k_2\sin(\eta) - \sqrt{k_1^2+k_2^2}\sin(\lambda t)},
$$
$$
v(\eta, t)=\frac{1}{\lambda}\cdot
\frac{\sqrt{k_1^2 - k_2^2}\cos(\lambda t)}
{k_1{\ch}(\eta) + k_2{\sh}(\eta) + \sqrt{k_1^2-k_2^2}\sin(\lambda t)},
$$
$$
v(\eta, t)=\frac{1}{\lambda}\cdot
\frac{\sqrt{k_1^2 - k_2^2}{\sh}(\lambda t)}
{k_1{\ch}(\eta) + k_2{\sh}(\eta) - \sqrt{k_1^2 - k_2^2}{\ch}(\lambda t)}.
$$
Here $\lambda\ne 0,  k_1>k_2$ are arbitrary numerical parameters.

{\bf Remark.}$\;$ The proof of an analog of Theorem 3 for the system of
quasilinear parabolic equations (7) has been described in the author's
paper [8].

It is known rather well [9, á.84] that the Laplace equation in the
2D coordinate space is invariant with respect to the conformal
transformation of independent variables
$$
\frac{\partial x}{\partial q_2}=\frac{\partial y}{\partial q_1},\;\;
\frac{\partial x}{\partial q_1}=-\frac{\partial y}{\partial q_2},
\eqno(14)
$$
furthermore, $\displaystyle\frac{\partial x}{\partial q_1}, \frac{\partial y}{\partial q_2}$
do not simultaneously turn zero. The system of parabolic coordinates
$$
x=q_2^2-q_1^2, \;\; y=2q_1q_2.
$$
is an example of such transformation.  Now, let us apply the
transformation (14) to eq. (1) under scrutiny. The latter may be
rewritten as follows:
$$
u_t=f(q_{1},q_{2})\Delta_{q_{1}q_{2}}\ln u, \;\; u=u(q_1,q_2,t).
\eqno(15)
$$
Here the denotation $f(q_1,q_2)=x_{q_1}^2+x_{q_2}^2$ is used.
Since due to (14) the function $x=x(q_1,q_2)$ is harmonic, Lemma 1
implies that
$$
\Delta_{q_{1}q_{2}}\ln f(q_{1},q_{2})=0.
\eqno(16)
$$
Consequently, the following theorem is valid.

{\bf Theorem 4.}$\;${\it
If relation (16) holds then with the aid of the transformation
$$
u(q_1,q_2,t)=\frac{1}{f(q_1,q_2)}\left [
f_{q_1}^2+f_{q_2}^2\right ]v(\eta,t),\;\;  \eta=\ln f(q_1,q_2),
\eqno(17)
$$
the nonhomogeneous equation of fast diffusion (15)
can be reduced to the homogenous and one-dimensional
(with respect to the spatial variable $\eta$) equation (12).
}

The proof of this theorem is similar to that of Theorem 3. To the end
of proving it is necessary to put $\eta (q_{1},q_{2})=\ln f(q_{1},q_{2})$
in (11) written in terms of variables $q_{1}$ and $q_{2}$.

{\bf Example 3.}$\;$  The equation of nonlinear diffusion
$$
u_t=\exp(x^2 - y^2)\Delta\ln u
$$
$$\hbox{or}\;\;\;
w_{t}=\exp(x^2 - y^2- w)\Delta w,
$$
where $w(x,y,t)=\ln v(x,y,t)$, has the following exact
solution, which is asymmetric with respect to the spatial
variables
$$
u(x,y,t)=4(x^2 + y^2)\exp (x^2 -y^2)v(\eta ,t), \;\; \eta=x^2 - y^2.
$$
Furthermore, the function $v(\eta ,t)$ satisfies eq. (12), some
exact solutions of which have been given in Example 2.

\section{Exact solutions of Liouville equation in the 2D
coordinate space}

It can easily be shown that principal results of the sections 1 and 2
can be directly generalized onto the equation of fast diffusion with
a linear source (sink)
$$
u_t=\Delta\ln u -\lambda u,
\eqno(18)
$$
where $\lambda\in \Re\setminus\{0\}.$ Since formulas (4),  (11),  (17)
are obviously independent of time, Theorems 1, 3 and 4 remain valid
for the steady-state (elliptical) eq. (18)
$$
\Delta\ln u=\lambda u,
$$
which by the substitution $u=\exp(\lambda w)$ may be reduced to
the well-known Liouville equation:
$$
\Delta w=\exp(\lambda w).
\eqno(19)
$$
So, Theorem 1 and Lemma 1 together imply the following

{\bf Proposition 1.}$\;${\it
The Liouville equation (19) is invariant with respect to the
transformation
$$
w(x,y)=v(\xi,\eta) + \frac{1}{\lambda}\ln |\rho (x,y)|,
\eqno(20)
$$
where $\rho =|\nabla\eta|^2,$ and $\xi (x,y),  \eta(x,y)$ are
conjugate harmonic functions.
}

By applying transformation (11) to eq. (19) it is possible to obtain
an ordinary differential equation (ODE), which can be easily integrated,
and make sure that the following proposition is valid.

{\bf Proposition 2.}$\;${\it
The Liouville equation (19) in the 2D coordinate space has exact
solutions of the form
$$
w(x,y)=\frac{1}{\lambda}\ln\left |
\frac{2A^2}{\lambda}\left (\eta_{x}^2 + \eta_{y}^2\right ){\sec}^2(A\eta)\right |,
$$
$$
w(x,y)=\frac{1}{\lambda}\ln\left |
\frac{2A^2}{\lambda}\left (\eta_{x}^2 + \eta_{y}^2\right ){\sech}^2(A\eta)\right |,
$$
where $\eta (x,y)$ is an arbitrary harmonic function, which is
different from the constant one, $A,\lambda\in\Re\setminus\{0\}.$
}

Similarly, from Theorem 4 it follows that for a steady-state
nonhomogeneous equation (15) having a linear
adjunct the following proposition is valid.

{\bf Proposition 3.}$\;${\it
The nonhomogeneous Liouville equation
$$
\Delta w = \eta (x,y)\exp(\lambda w),
\eqno(21)
$$
where $\eta (x,y)$ is an arbitrary harmonic function different from
the constant one, has the exact solution
$$
w(x,y)=\frac{1}{\lambda}\ln\left |\frac{3}{\lambda}
\frac{\eta_{x}^2 + \eta_{y}^2}{\eta^3}\right |.
$$
}

Note that exact solutions of the nonhomogeneous
Liouville equation (21), when $\eta (x,y)$ is a holomorphic
function of special form, can be found in [10].

As a case of generalization of the results obtained in this section onto
some different systems of equations consider the steady-state mathematical
model of charge transfer, which implies Poisson interactions,
known from the theory of semiconductors.

{\bf Example 4.}$\;$ The elliptical system of equations of the form
$$
\Delta u=\exp(u) + A|\nabla "|^2,\;\;
\Delta v=\exp(v) - B|\nabla "|^2,
$$
$$
\Delta "=\exp(v) - \exp(u), \eqno(22)
$$
has a exact solution of the form
$$
u(x,y)=f(\eta) + \ln\left|\eta_{x}^2 + \eta_{y}^2\right|,\;
v(x,y)=\psi (\eta) + \ln\left|\eta_{x}^2 + \eta_{y}^2\right|,\;
"(x,y)=\varphi(\eta),
$$
where $\eta (x,y)$ is an arbitrary harmonic function, which is
different from the constant one; $ A, B\in\Re\setminus\{0\}.$
Furthermore, the functions $f, \psi, \varphi$ can be defined from
the system of ODE
$$
f^{\prime\prime}=\exp(f) + A\varphi^{\prime^2},\;
\psi^{\prime\prime}=\exp(\psi) - B\varphi^{\prime^2},\;
\varphi^{\prime\prime}=\exp(\psi) - \exp(f).
$$
In the case when $A=-B,$ system (22) assumes a particular exact
solution
$u(x,y)=v(x,y)=f(\eta) + \ln\left|\eta_{x}^2 + \eta_{y}^2\right|,
"(x,y)=\eta (x,y),$
furthermore, the function $f(\eta)$ satisfies the linear 2nd order ODE
$$
f^{\prime\prime}=f + A.
$$
\section*{\bf Acknowledgement}

This work was supported by an INTAS grant (No. 2000-15).

\end{document}